\documentclass[fleqn,usenatbib,useAMS]{mnras}

\usepackage{newtxtext,newtxmath}

\usepackage[T1]{fontenc}

\DeclareRobustCommand{\VAN}[3]{#2}
\let\VANthebibliography\thebibliography
\def\thebibliography{\DeclareRobustCommand{\VAN}[3]{##3}\VANthebibliography}

\usepackage{graphicx}	
\usepackage{amsmath}	
\usepackage{orcidlink}
\usepackage{xspace}

\newcommand{\msun}{{\rm M}_{\sun}}
\newcommand{\phone}{WD 2359$-$434\xspace}
\newcommand{\another}{WD 2105$-$820\xspace}

\title[White dwarf surface dynamos]{Magnetic dynamos powered by white dwarf superficial convection}

\author[R. Yaakovyan et al.]{
Rom Yaakovyan$^{\orcidlink{0009-0000-7852-8815}}$,$^{1}$\thanks{\raggedright E-mail: \href{mailto:rom.yaakovyan@mail.huji.ac.il}{rom.yaakovyan@mail.huji.ac.il} (RY);
\href{mailto:sivan.ginzburg@mail.huji.ac.il}{sivan.ginzburg@mail.huji.ac.il} (SG)}\label{emails}
Sivan Ginzburg$^{\orcidlink{0000-0002-3751-4553}}$,$^{1}$\hyperref[emails]{\footnotemark[1]} Jim Fuller$^{\orcidlink{0000-0002-4544-0750}}$$^{2}$ and Nicholas Z. Rui$^{\orcidlink{0000-0002-1884-3992}}$$^{2}$
\\
$^{1}$Racah Institute of Physics, The Hebrew University, Jerusalem 9190401, Israel\\
$^{2}$TAPIR, California Institute of Technology, Pasadena, CA 91125, USA
}

\date{Accepted XXX. Received YYY; in original form ZZZ}

\pubyear{\the\year{}}

\begin{document}
\label{firstpage}
\pagerange{\pageref{firstpage}--\pageref{lastpage}}
\maketitle

\begin{abstract}
When the effective temperature of a cooling white dwarf $T_{\rm eff}$ drops below the ionization limit, it develops a surface convection zone that may generate a magnetic field $B$ through one of several dynamo mechanisms. We revisit this possibility systematically using detailed stellar evolution computations, as well as a simple analytical model that tracks the expansion of the convection zone. The magnetic field reaches a maximum of several kG (for a hydrogen atmosphere) shortly after a convection zone is established at a cooling time $t=t_{\rm conv}$. The field then declines as $B\propto T_{\rm eff}\propto t^{-7/20}$ until the convective envelope couples to the degenerate core at $t=t_{\rm coup}$. We compare the onset of convection $t_{\rm conv}\propto M^{25/21}$ to the crystallization of the white dwarf's core $t_{\rm cryst}\propto M^{-5/3}$, and find that in the mass range $0.5\,\msun<M<0.9\,\msun$ the order of events is $t_{\rm conv}<t_{\rm cryst}<t_{\rm coup}$. Specifically, surface dynamos are active for a period $\Delta t\approx t_{\rm cryst}-t_{\rm conv}$ of about a Gyr (shorter for higher masses), before the convection zone is overrun by a stronger magnetic field emanating from the crystallizing core. Our predicted magnetic fields are at the current detection limit, and we do not find any observed candidates that fit the theory. None the less, surface dynamos may be an inevitable outcome of white dwarf cooling, significantly affecting white dwarf accretion and seismology.  
\end{abstract}




\section{Introduction}

Observations indicate that a large fraction of white dwarf stars are magnetic, with a potentially bimodal population of weak ($\lesssim 10^5$ G) and strong ($10^6-10^9$ G) magnetic fields \citep[see][for reviews]{Ferrario2015,Ferrario2020}. These magnetic fields may have been inherited from a previous stage of stellar evolution \citep[i.e. the fossil field hypothesis; see][]{LevyRose1974,Angel1981,BraithwaiteSpruit2004,Tout2004,WickramasingheFerrario2005}.
Alternatively, magnetic fields might be the result of a double white dwarf merger \citep{GarciaBerro2012} or a common envelope event \citep{RegosTout1995,Tout2008,Nordhaus2011}. 

More recently, \citet{Isern2017} postulated that white dwarfs may generate magnetic fields as part of their standard cooling process, through a convective dynamo operating during their crystallization -- when their cores gradually solidify from the inside out. Because white dwarfs begin crystallizing at a cooling age of several Gyr (depending on their mass), crystallization dynamos are a promising mechanism to explain the late appearance of strong magnetic fields in volume-limited samples \citep{BagnuloLandstreet2022}. However, both the magnitude and the exact timing of crystallization dynamos remain uncertain \citep{Schreiber2021Nat,Ginzburg2022,Fuentes2023,Fuentes2024,BlatmanGinzburg2024,Castro-Tapia2024,Castro-Tapia2024b,MontgomeryDunlap2024}, and the late appearance of strong magnetism may be attributed instead to the diffusion of buried fields generated during the main sequence or giant phase to the white dwarf's surface \citep{Camisassa2024}.  

Here, we focus on another magnetic dynamo operating during white dwarf cooling. As the effective temperature of a white dwarf decreases, a recombination front appears at the surface and gradually advances inward. The partial ionization above this front changes the opacity, forming a surface convection zone that expands inward as the white dwarf continues to cool to lower temperatures \citep[e.g.][]{Bohm1968,vanHorn1970,Fontaine2001}. Dynamos powered by the surface convection zone -- combined with the white dwarf's rotation -- have been considered in the past as a source of white dwarf magnetism. 
Specifically, several dynamo mechanisms have been suggested:
\begin{enumerate}
    \item 
    The $\alpha\Omega$ dynamo, which requires convection and differential rotation \citep{Parker1955,SteenbeckKrause1969a,Fontaine1973,Markiel1994,Thomas1995}. 
    Asteroseismology provides evidence for differential rotation in some white dwarfs \citep{Winget1994,Kawaler1999,Corsico2011,Corsico2022} but not in others \citep{Charpinet2009,Giammichele2016,Giammichele2018}. In any case, the extent of differential rotation in the thin convective envelope remains unclear.    
    \item 
    The $\alpha^2$ dynamo, which requires convection and only uniform rotation \citep{SteenbeckKrause1969b,Fontaine1973,Raedler1980,Brandenburg2017}.
    \item 
    A local turbulent dynamo, which requires only convection \citep{Cattaneo1999,VolgerSchussler2007,Moll2011}. Unlike the global $\alpha\Omega$ and $\alpha^2$ dynamos, where rotation may enforce a large-scale magnetic field (such as a dipole or a quadrupole), this dynamo produces small-scale magnetic fields, limited by the size of the convective eddies. As discussed in Section \ref{sec:obs}, this may hinder their observational detection \citep[see also][]{Tremblay2015}. 
\end{enumerate}

Determining whether white dwarfs satisfy the conditions for any of these dynamos to operate is beyond the scope of this work. Instead, we rely on equipartition arguments between the magnetic and kinetic energies in the convective eddies, which limit the magnetic field to $\lesssim 10^4$ G for a hydrogen atmosphere, regardless of the dynamo's nature \citep{Fontaine1973,Markiel1994,Thomas1995,Christensen2009,Tremblay2015}.

Although surface convection dynamos cannot account for the strongly magnetized white dwarfs, they may still be an unavoidable outcome of white dwarf cooling, at least for some rotation rates. 
Motivated by recent observations that have pushed the detection limit to a few kG \citep{AznarCuadrado2004,Jordan2007,KawkaVennes2012,Landstreet2012,Landstreet2015,BagnuloLandsreet2018,BagnuloLandstreet2019,BagnuloLandstreet2022,Farihi2018} and by the potential effect of such weak fields on white dwarf seismology \citep{Rui2025} and accretion \citep{Metzger2012,HarringtonGaraud2019,Fraser2024}, we revisit here the surface dynamo theory systematically in the context of other events during white dwarf cooling, such as crystallization and the coupling of the surface convection zone to the degenerate core \citep{Fontaine2001}. 

The remainder of this paper is organized as follows. In Section \ref{sec:anal} we develop a simple analytical model for the white dwarf's convection zone and magnetic dynamo. In Section \ref{sec:numr} we evolve white dwarfs numerically using the \textsc{mesa} stellar evolution code, and compare these detailed calculations to our analytical scaling relations. We compare our theory to the observations in Section \ref{sec:obs}, and summarize our conclusions in Section \ref{sec:conclusions}.

\section{Analytical model}\label{sec:anal}

\subsection{Convection zone}

To provide intuition and context for our numerical results, we begin with a simple analytical model for the surface convection zone, based on the classical work of \cite{Mestel1952}, which we briefly repeat here for completeness \citep[see][for more details]{vanHorn1971}. In this model, the white dwarf consists of a degenerate core with a mass $M$ and a radius $R\propto M^{-1/3}$, which is roughly isothermal at a temperature $T=T_{\rm c}$ thanks to the high conductivity of degenerate electrons. As the density decreases towards the white dwarf's edge, the core transitions into an ideal gas envelope (with negligible mass and thickness) that regulates its cooling. This transition from degenerate pressure $P\propto\rho^{5/3}$ to ideal gas $P\propto\rho T$ occurs at a density $\rho_{\rm deg}\propto T_{\rm c}^{3/2}$ and pressure $P_{\rm deg}\propto T_{\rm c}^{5/2}$.
Assuming that heat is transported trough the ideal gas envelope by photon diffusion and adopting Kramers' opacity $\kappa\propto\rho T^{-7/2}$, the optical depth $\tau\sim \kappa P/g$ to the degenerate core scales as $\tau_{\rm deg}\propto T_{\rm c}^{1/2}/g\propto T_{\rm c}^{1/2}R^2/M$, where $g$ is the surface gravity. Using the diffusion equation, the photon flux is given by $F\propto {\rm d}T^4/{\rm d}\tau\sim T_c^4/\tau_{\rm deg}$. The white dwarf's luminosity is therefore $L\sim R^2F\propto R^2T_{\rm c}^4/\tau_{\rm deg}\propto MT_{\rm c}^{7/2}$ and the cooling time $t\sim E_{\rm th}/L\propto MT_{\rm c}/L\propto T_{\rm c}^{-5/2}$, where $E_{\rm th}$ is the white dwarf's thermal energy.

As the white dwarf cools down, its flux drops over time as
\begin{equation}\label{eq:flux}
    F\sim\frac{L}{R^2}\propto M^{5/3}T_{\rm c}^{7/2}\propto M^{5/3}t^{-7/5},
\end{equation}
and the effective temperature decreases as
\begin{equation}\label{eq:teff}
    T_{\rm eff}\propto F^{1/4}\propto M^{5/12}t^{-7/20}.
\end{equation}
At some point, $T_{\rm eff}$ (i.e. the temperature at an optical depth $\tau\sim 1$) drops below the ionization temperature $T_0$, leading to partial recombination of the white dwarf's outer layers. This changes the opacity, rendering the temperature profile unstable to convection \citep{vanHorn1970,Fontaine2001}. For simplicity, we approximate $T_0\approx{\rm const.}$, neglecting the dependence of ionization on the density and squeezing the recombination front into a sharp temperature threshold. With this approximation (which explains well our main numerical results), convection first appears at a cooling time
\begin{equation}\label{eq:t_conv}
    t_{\rm conv}\propto\left(\frac{M^{5/3}}{T_0^4}\right)^{5/7}\propto M^{25/21}.
\end{equation}

Initially, the appearance of a surface convection zone does not alter the white dwarf's cooling rate. The degenerate core remains insulated by an optically thick radiative mantle that acts as a thermal bottleneck. In fact, the optical depth is dominated by the bottom of the radiative layer (i.e. at the degeneracy boundary), such that the cooling model derived  by \citet{Mestel1952} remains applicable \citep{Fontaine2001,Tremblay2015}.\footnote{As outlined by \citet{vanHorn1971} and \citet{Fontaine2001}, the classical \citet{Mestel1952} model suffers from many inaccuracies and misses key physical processes, such as delayed cooling during crystallization and accelerated Debye cooling. None the less, it serves as a useful analytical starting point.} As the white dwarf continues to cool and more layers recombine, the convection zone expands inward to higher densities. Inside the radiative region, the diffusion equation dictates $F\propto T^4/\tau$. Specifically, at the radiative--convective boundary (i.e. the bottom of the convection zone) the optical depth scales as
\begin{equation}\label{eq:tau_conv}
    \tau_{\rm conv}\propto \frac{T_0^4}{F}\propto M^{-5/3}t^{7/5},
\end{equation}
where the flux is given by equation \eqref{eq:flux}. At $t=t_{\rm conv}$, $\tau_{\rm conv}\sim 1$, such that photon diffusion reaches all the way to the photosphere and there is no convection. Using hydrostatic equilibrium and Kramers' opacity law
\begin{equation}\label{eq:tau_conv_rho}
    \tau_{\rm conv}\sim\frac{\kappa P_{
    \rm conv}
    }{g}\propto\frac{R^2\rho_{\rm conv}^2T_0^{-5/2}}{M}\propto M^{-5/3}\rho_{\rm conv}^2, 
\end{equation}
where $\rho_{\rm conv}$ and $P_{\rm conv}\propto\rho_{\rm conv}T_0$ are the density and pressure at the bottom of the convection zone. By comparing equations \eqref{eq:tau_conv} and \eqref{eq:tau_conv_rho} we find that the convection zone expands as
\begin{equation}\label{eq:rho_conv}
    \rho_{\rm conv}\propto t^{7/10}.
\end{equation}

As the convection zone expands from the surface to higher densities, the degenerate core expands to lower densities $\rho_{\rm deg}\propto T_{\rm c}^{3/2}\propto t^{-3/5}$ until $\rho_{\rm conv}(t)\sim\rho_{\rm deg}(t)$ and the convective envelope directly couples to the degenerate core -- eliminating the radiative bottleneck that separated them. From this convective-coupling time $t=t_{\rm coup}\propto M^0$ onward, the \citet{Mestel1952} model is no longer appropriate and the cooling rate changes \citep{Fontaine2001,Tremblay2015,Ginzburg2024}. Our analytical model is therefore limited to times $t_{\rm conv}<t<t_{\rm coup}$. As we show in Section \ref{sec:numr}, this is the relevant regime for surface dynamos in most of the white dwarf mass range.

\subsection{Magnetic field}

In the convection zone, the flux $F$ is carried by fluid motion with a velocity $v_{\rm conv}$ that is given by standard mixing-length theory
\begin{equation}\label{eq:v_conv}
    F\sim \rho v_{\rm conv}^3.
\end{equation}
This motion of the electrically conducting fluid is thought to induce a magnetic field $B$ through a dynamo mechanism. Regardless of the exact nature of the dynamo, it is generally thought that the field's strength is limited by equipartition between the kinetic and magnetic energy densities \citep{Fontaine1973,Tremblay2015}
\begin{equation}\label{eq:equipartition}
    \frac{B^2}{8\upi}\sim\frac{1}{2}\rho v_{\rm conv}^2,
\end{equation}
although stronger fields might be possible if the white dwarf rotates sufficiently fast \citep[low Rossby numbers; see][]{Augustson2016,Augustson2019}. In our case, the typical convective turnover times $\sim\Delta r_{\rm conv}/v_{\rm conv}$ are measured in hours, such that only the fastest spinning white dwarfs \citep{Hermes2017,OliveiradaRosa2024} have Rossby numbers $\textrm{Ro}\lesssim 1$.\footnote{In our analytical model, and also numerically (except for a short transient at the onset of convection), the depth of the convection zone is set by the scale height at its base $\Delta r_{\rm conv}\propto T_0/g$, which remains approximately constant (because $T_0\approx\textrm{const.}$) even as $\rho_{\rm conv}$ increases. The convective velocity, on the other hand, decreases as $v_{\rm conv}\sim(F/\rho_{\rm conv})^{1/3} \propto t^{-7/10}$.}

By combining equations \eqref{eq:v_conv} and \eqref{eq:equipartition}, the magnetic field in the convection zone is given by
\begin{equation}\label{eq:B_rho}
    B\sim \rho^{1/2}\left(\frac{F}{\rho}\right)^{1/3}\sim \rho^{1/6} F^{1/3}.
\end{equation}
Because the flux is uniform in the white dwarf's outer layers, the strongest magnetic field is produced at the bottom of the convection zone (where $\rho=\rho_{\rm conv}$), and using equations \eqref{eq:flux}  and \eqref{eq:rho_conv} it scales as
\begin{equation}\label{eq:B_overall}
    B\sim\rho_{\rm conv}^{1/6}F^{1/3}\propto t^{7/60}M^{5/9}t^{-7/15}\propto M^{5/9}t^{-7/20}.
\end{equation}
Equation \eqref{eq:B_overall} shows that the increasing density of the convection zone competes against the decreasing flux, such that the white dwarf's magnetic field decreases over time from a maximal
\begin{equation}
    B_{\rm max}\propto M^{5/9}t_{\rm conv}^{-7/20}\propto M^{5/36}
\end{equation}
at $t=t_{\rm conv}$, which is given by equation \eqref{eq:t_conv}.
Alternatively, using equations \eqref{eq:teff} and \eqref{eq:B_overall} we may write $B(T_{\rm eff})$
\begin{equation}\label{eq:B_Teff}
    B\propto M^{5/36}T_{\rm eff},
\end{equation}
with $B_{\rm max}$ obtained at the onset of convection, when $T_{\rm eff}=T_0$. Equation \eqref{eq:B_Teff} provides a more accurate scaling relation than \citet{Fontaine1973}, who neglected the expansion of the convection zone and derived a somewhat steeper decrease as the white dwarf cools down $B\propto T_{\rm eff}^{4/3}$.

\section{Numerical results}\label{sec:numr}

We used the stellar evolution code \textsc{mesa} \citep{Paxton2011,Paxton2013,Paxton2015,Paxton2018,Paxton2019,Jermyn2023}, version r24.08.1, to create DA (hydrogen-atmosphere) carbon--oxygen white dwarf models in the mass range $0.5\,\msun\leq M \leq 0.9\,\msun$. We note that ionization and the surface convection dynamo are very sensitive to the composition of the atmosphere \citep{Fontaine1973}, but we focus here mainly on DA white dwarfs, which comprise the majority of the relevant observations (Section \ref{sec:obs}). Preliminary computations of DB (helium-atmosphere) white dwarfs are briefly discussed in Appendix \ref{sec:helium} for completeness. 
The white dwarf models are evolved from pre-main-sequence progenitors with different initial masses using the test suite \texttt{make\_co\_wd}, and are then cooled down using the test suite \texttt{wd\_cool\_0.6M}.

\begin{figure}
    \centering
    \includegraphics[width=1\linewidth]{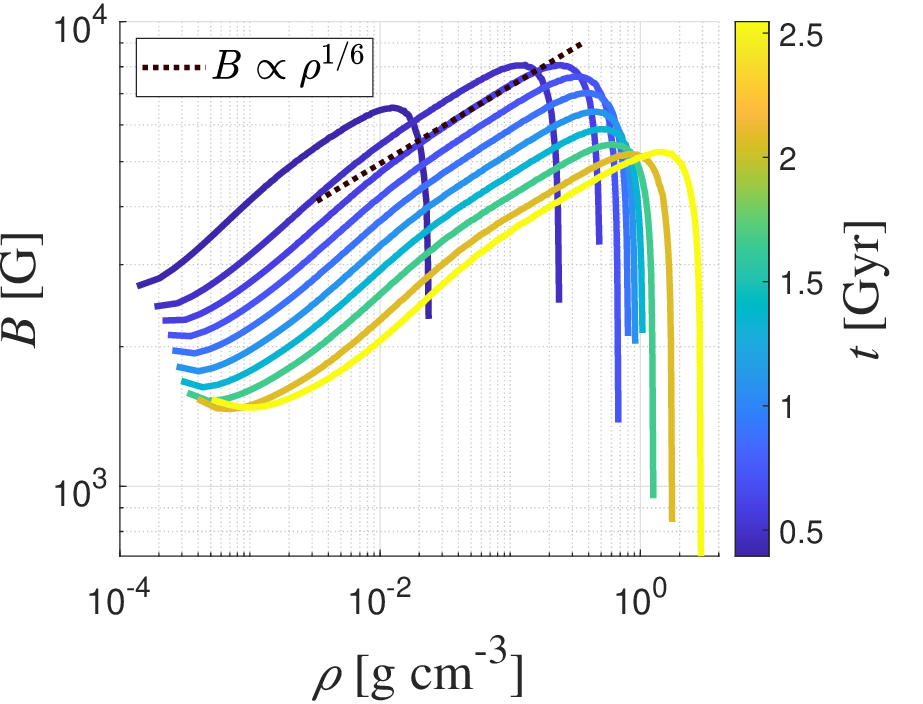}
    \caption{The magnetic field $B$ profile as a function of density $\rho$ inside a $0.6\,\msun$ white dwarf at different (colour coded) cooling times $t$, between the onset of the convection and the convective-coupling time ($t_{\rm conv}<t<t_{\rm coup}$). The approximately uniform flux through the white dwarf's convective envelope implies $B\propto\rho^{1/6}$ according to equation \eqref{eq:B_rho}.}
    \label{fig:B_rho}
\end{figure}

\begin{figure}
    \centering
    \includegraphics[width=1\linewidth]{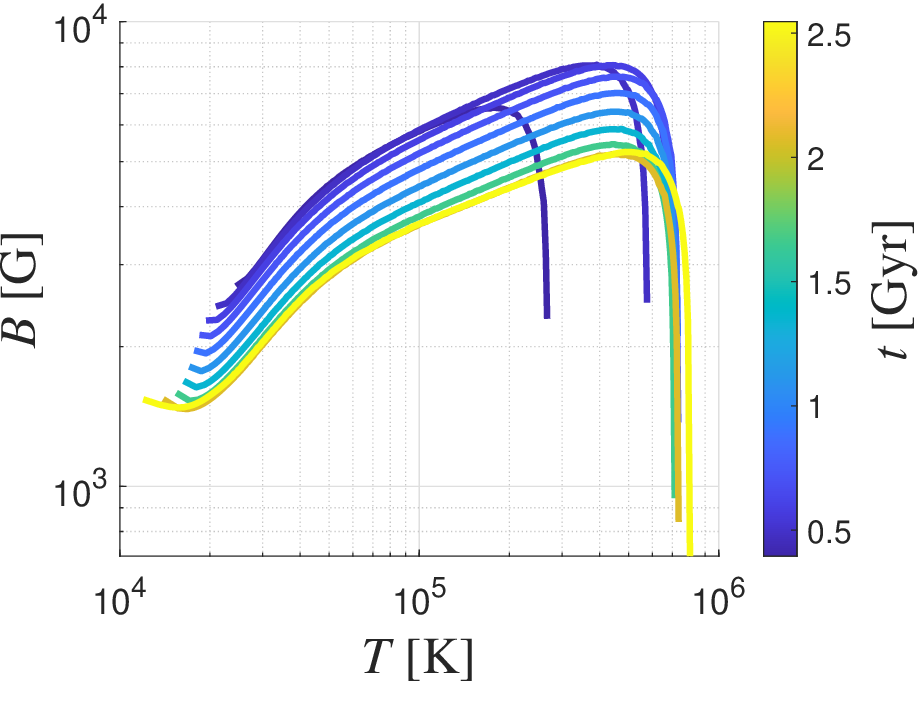}
    \caption{Same as Fig. \ref{fig:B_rho}, but as a function of temperature $T$.}
    \label{fig:B_T}
\end{figure}

At any given time, we compute the magnetic field $B$ in each convective cell ($v_{\rm conv}>0$) in the simulation using equation \eqref{eq:equipartition}. As seen in Fig. \ref{fig:B_rho}, the strongest magnetic field is generated close to the bottom of the convection zone, which gradually advances to higher pressures and densities as the white dwarf cools down. After a short transient, the temperature at the bottom of the convection zone saturates at a roughly constant value, as seen in Fig. \ref{fig:B_T}, justifying our analytical scaling relations.

\begin{figure}
    \centering
    \includegraphics[width=1\linewidth]{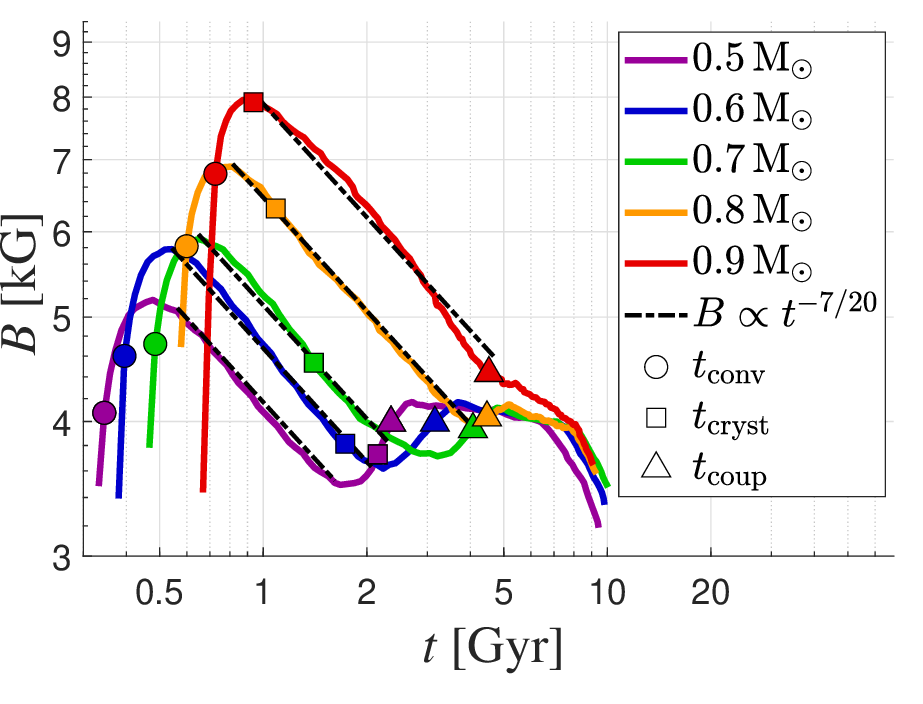}
    \caption{The maximal magnetic field $B$ as a function of the white dwarf's cooling time $t$ for different white dwarf masses. Markers indicate the establishment of a convection zone $t_{\rm conv}$, the core crystallization time $t_{\rm cryst}$, and the convective-coupling time $t_{\rm coup}$. During times $t_{\rm conv}\lesssim t\lesssim t_{\rm coup}$, the field's evolution is described well by equation \eqref{eq:B_overall}: $B\propto t^{-7/20}$.}
    \label{fig:B_time}
\end{figure}

\begin{figure}
    \centering
    \includegraphics[width=1\linewidth]{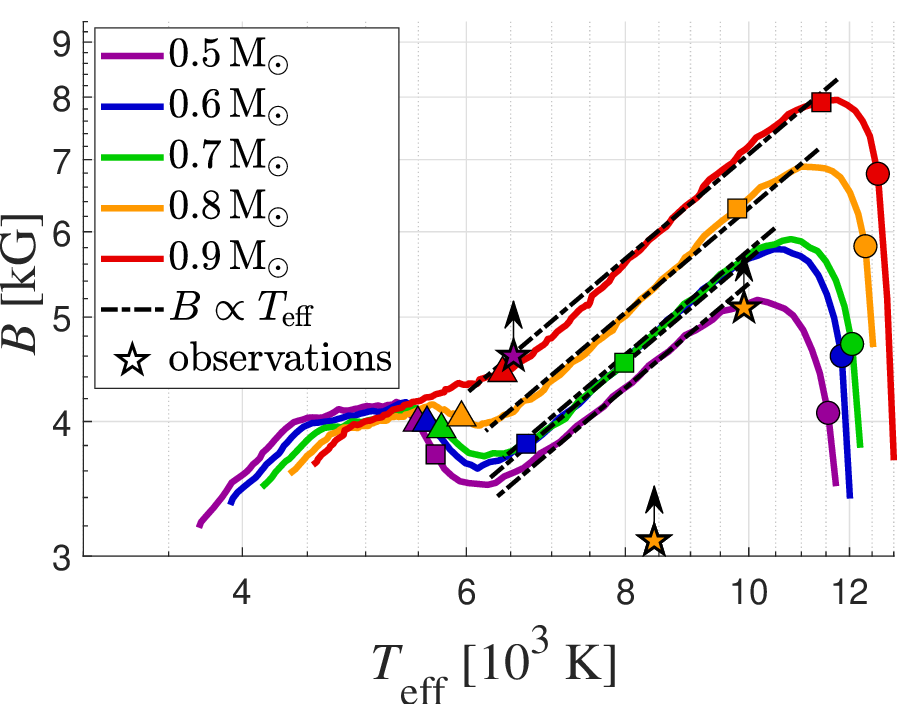}
    \caption{Same as Fig. \ref{fig:B_time}, but as a function of the effective temperature $T_{\rm eff}$. The field's decline during $t_{\rm conv}\lesssim t\lesssim t_{\rm coup}$ is described well by equation \eqref{eq:B_Teff}: $B\propto T_{\rm eff}$. We identify three white dwarfs in the literature that are cold enough and have sufficiently low kG magnetic field measurements to be explained by a surface convection dynamo (see also Fig. \ref{fig:time_points}). Star markers (coloured as the model with the closest mass) indicate their measured longitudinal component $B_z$, which sets a lower limit on the total field $B$. Additional observations indicate that one star is a helium-core white dwarf, while the other two likely have a much stronger total magnetic field $B\gg B_z$ (see text). None the less, the figure demonstrates that kG magnetic fields produced by surface dynamos are within the reach of spectropolarimetry.}
    \label{fig:B_Teff}
\end{figure}

In Figs \ref{fig:B_time} and \ref{fig:B_Teff} we plot the maximal magnetic field $B$ as a function of time $t$ and effective temperature $T_{\rm eff}$. After a short transient at the onset of convection, the numerical results fit well our analytical power laws from Section \ref{sec:anal}: $B\propto t^{-7/20}$ and $B\propto T_{\rm eff}$. These power laws break down close to the convective-coupling time $t_{\rm coup}$, when the convection zone penetrates into the degenerate interior, defined here by the degeneracy parameter $\eta>0$ \citep{Fontaine2001,Tremblay2015}.\footnote{$\eta kT$ is the electron chemical potential ($k$ is the Boltzmann constant), which is positive for degenerate electrons and negative for an ideal gas \citep{Kippenhahn2012}.} The dependence on the mass $B(M)$ is somewhat steeper than predicted in Section \ref{sec:anal}, partially due to inaccuracies of the underlying \citet{Mestel1952} cooling model. Fig. \ref{fig:B_Teff} demonstrates that a convection zone first appears at roughly the same $T_{\rm eff}$ for all masses, justifying our analytical scaling relations. However, this $T_{\rm eff}$ differs from the saturated temperature at the base of the convection zone (see Fig. \ref{fig:B_T}), unlike in our simplistic analytical model -- where both temperatures equal the same constant $T_0$.

\begin{figure}
    \centering
    \includegraphics[width=1\linewidth]{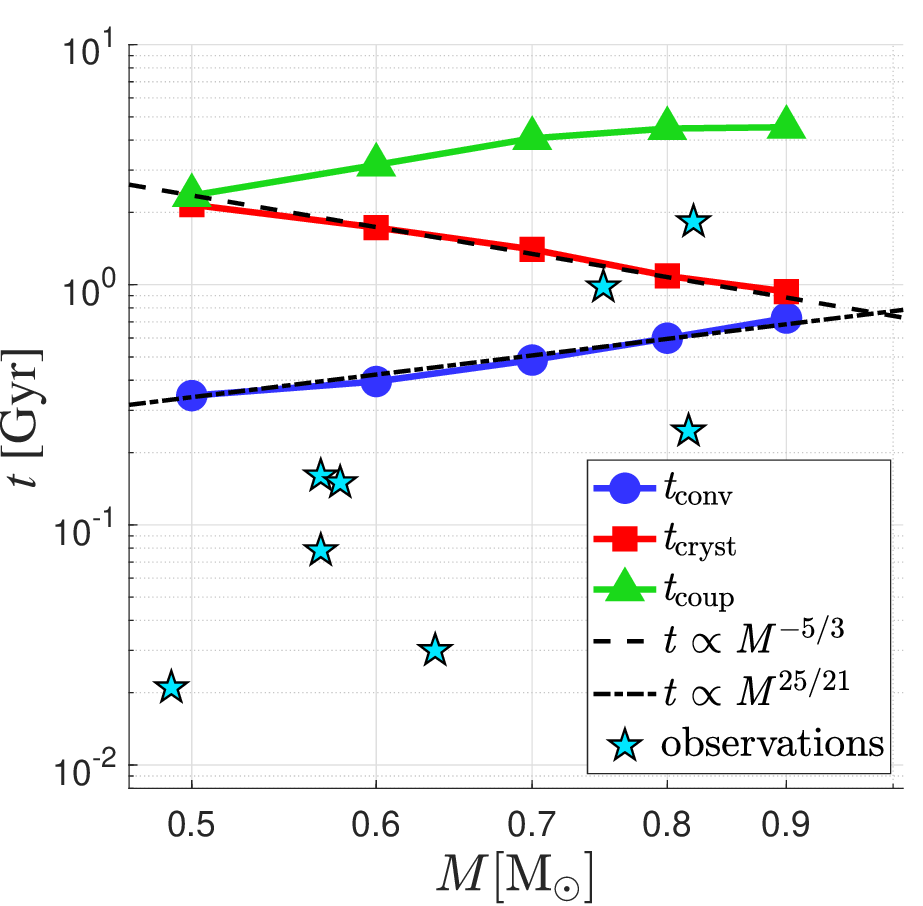}
    \caption{The onset of convection $t_{\rm conv}$, crystallization $t_{\rm cryst}$, and convective-coupling $t_{\rm coup}$ times in carbon--oxygen white dwarfs with different masses $M$. The convection time fits equation \eqref{eq:t_conv}: $t_{\rm conv}\propto M^{25/21}$, whereas the crystallization time fits $t_{\rm cryst}\propto M^{-5/3}$ \citep{BlatmanGinzburg2024}. The convective-coupling time deviates from our analytical estimate $t_{\rm coup}\propto M^0$; see the discussion below equation \eqref{eq:rho_conv}. 
    For the entire mass range, $t_{\rm conv}<t_{\rm cryst}<t_{\rm coup}$. Only two of the  carbon--oxygen white dwarfs with kG magnetic field measurements (star markers) are old enough to be explained by a surface convection dynamo. Another low-mass helium white dwarf cools at a different pace $t(T_{\rm eff})$, and is therefore not shown in this figure (but appears in Fig. \ref{fig:B_Teff}, because convection starts at approximately the same $T_{\rm eff}$ regardless of the core's composition).}
    \label{fig:time_points}
\end{figure}

In Fig. \ref{fig:time_points} we plot relevant time points in the evolution of white dwarfs as a function of their mass. We determine the onset of convection $t_{\rm conv}$ by the time when a substantial convection zone has formed -- penetrating into densities $\rho\gtrsim 10^{-2}\textrm{ g cm}^{-3}$. As seen in Fig. \ref{fig:B_time}, our results are insensitive to this arbitrary choice thanks to the fast rise of the dynamo field to its peak value. After the convective-coupling time $t_{\rm coup}$ (defined above), our analytical model breaks down, but even before that the weak field generated by a surface dynamo may be overrun by that of a deeper crystallization dynamo, depending on its strength.
We estimate the onset of crystallization in the white dwarf's core $t_{\rm cryst}$ as the time when the plasma coupling parameter $\Gamma>200$ at the centre. While the exact value of the critical $\Gamma$ depends on the carbon to oxygen ratio \citep{Jermyn2021}, this approximation is sufficient for our purposes. Fig. \ref{fig:time_points} shows that for our entire mass range $t_{\rm conv}<t_{\rm cryst}<t_{\rm coup}$, such that a surface dynamo always operates for some period of time, before it is overrun by the crystallization dynamo; our simple analytical model is applicable for most of this period. 

\section{Comparison to observations}\label{sec:obs}

In Figs \ref{fig:B_Teff} and \ref{fig:time_points} we compare our theoretical predictions to the population of observed white dwarfs with measured magnetic fields $B<10^4\textrm{ G}$ (all of spectral type DA, i.e. hydrogen atmospheres). This sample is composed of 5 stars from \citet{AznarCuadrado2004}, \citet{Landstreet2012}, and \citet{BagnuloLandstreet2022} with weak magnetic fields detected at the $3\sigma$ uncertainty level (WD 0232$+$525, WD 0446$-$789, WD 1105$-$048, WD 2105$-$820, WD 2359$-$434), as well as 3 stars from \citet{Jordan2007} with $2.4\sigma$ detections (WD 1620$-$391, WD 2007$-$303, WD 2039$-$202). An additional weakly magnetized white dwarf (NLTT 347) was identified by \citet{KawkaVennes2012}, but recent \textit{Gaia} measurements place its mass firmly in the low-mass helium white dwarf range \citep[$M\approx 0.34\,\msun$,][]{Caron2023,Vincent2024}. None the less, we plot it in Fig. \ref{fig:B_Teff} for completeness.
The kG magnetic fields in all of these stars were detected using spectropolarimetry, which measures the longitudinal component of the magnetic field $B_z$, hence providing a lower limit to the total field $B>B_z$. 

From this sample, we identify two candidates for carbon--oxygen white dwarfs 
magnetized by a surface convection dynamo -- \another ($M\approx 0.75\,\msun$, $T_{\rm eff}\approx 9900\textrm{ K}$) and
\phone ($M\approx 0.82\,\msun$, $T_{\rm eff}\approx 8400\textrm{ K}$), with the masses and effective temperatures given by \citet{O'Brien2024}. The other stars have not developed a substantial convection zone yet, requiring a different origin for their magnetic fields. The $B_z=3.1\textrm{ kG}$ measurement for \phone \citep{AznarCuadrado2004} and the $B_z=5.1\textrm{ kG}$ measurement for \another \citep{Farihi2018} are consistent with our $B(M,T_{\rm eff})\approx 5-6\textrm{ kG}$ theoretical estimates. Fig. \ref{fig:time_points} indicates that $t<t_{\rm cryst}$ for \another, whereas
\phone has already started crystallizing such that $t>t_{\rm cryst}$. However, \citet{BlatmanGinzburg2024} find that the magnetic field from a crystallization dynamo is initially buried deep inside the white dwarf's interior. It diffuses to the outer layers and eventually breaks out at the surface only at a cooling age $t_{\rm break}\approx 3\textrm{ Gyr}$ for a $0.8\,\msun$ white dwarf, such that for \phone, $t_{\rm cryst}<t<t_{\rm break}$. This is also true empirically, regardless of the theoretical origin of the magnetic field, as seen by the appearance time of strong magnetism in \citealt{BagnuloLandstreet2022}. We conclude that the ages of both our candidates are consistent with a surface convection dynamo, because the
convection zone has not been overrun yet by a magnetic field emanating from the core. Furthermore, we hypothesize that an active surface dynamo could have likely erased any pre-existing weaker fossil magnetic field. 

Alas, after the initial $B_z$ measurement of \phone by \citet{AznarCuadrado2004}, a much stronger total $B\sim 50-100 \textrm{ kG}$ was inferred from higher-resolution spectroscopy \citep{Koester2009,Landstreet2017}. The small $B_z/B\ll 1$ ratio is difficult to explain, requiring a very large inclination angle and a complex magnetic field geometry \citep{Landstreet2017}.\footnote{We note that dynamos that rely on rotation may naturally generate magnetic fields that are very different in the radial and azimuthal directions \citep{Spruit2002,Fuller2019}.} 
Alternatively, this small ratio may indicate a small-scale field structure: polarimetry is sensitive to the integrated $B_z$ over the visible stellar surface, which may change its polarity on a small scale. Spectroscopy, on the other hand, is sensitive to the magnitude $B$, which adds up coherently regardless of the field's orientation.
Despite these difficulties in modelling \phone's magnetic field, we conclude that this candidate for a surface convection dynamo is probably ruled out. Similarly, our second candidate \another also has a measured total $B\approx 43\textrm{ kG}$ using intensity spectroscopy \citep{Koester1998, Koester2009}. None the less, these white dwarfs demonstrate that spectropolarimetry is capable of detecting surface dynamos, unless they themselves have a small-scale structure.  

Seismology of ZZ Ceti (DAV)\footnote{These have hydrogen atmospheres. DBV stars are similar pulsating white dwarfs with helium atmospheres.} stars may offer an alternative path to measure magnetic fields generated by a surface dynamo. The pulsations of these stars are very sensitive to magnetic fields in the radiative layers just beneath the surface convection zone. Recently, \citet{Rui2025} used observed pulsations to estimate upper limits to near-surface magnetic fields, which can be less than a kG in some cases, below our predictions. However, it is unclear whether a magnetic field generated in the surface convection zone can penetrate (e.g. by magnetic diffusion or convective overshoot) into the radiative interior within which the pulsations propagate. Additionally, the narrow ZZ Ceti instability strip occurs when the surface convection zone is very shallow and the dynamo is first activated \citep[][]{FontaineBassard2008,WingetKepler2008}. The interplay between the surface dynamo and white dwarf pulsations should be investigated more thoroughly by focusing on this relatively short transient period (our current paper focuses instead on the much longer decline of the magnetic field at later times; see Figs \ref{fig:B_time} and \ref{fig:B_Teff}).   

\section{Conclusions}\label{sec:conclusions}

The strong MG magnetic fields observed in many white dwarfs have been the focus of intense research. Observationally, volume-limited samples indicate that strong magnetism appears late in the cooling phase of white dwarfs \citep{BagnuloLandstreet2022}. Theoretically, the crystallization dynamo \citep{Isern2017} offers a potential explanation for this delay, though other possibilities exist \citep{Camisassa2024}.
Here, we focused instead on the much weaker kG fields that are at the limit of spectropolarimetry detection capabilities, and have been so far measured only for a handful of white dwarfs \citep{AznarCuadrado2004,Jordan2007,KawkaVennes2012,Farihi2018,BagnuloLandstreet2022}. Such weak fields may impact accreting white dwarfs \citep{Metzger2012,Farihi2018,Cunningham2021}, and might also be detected by seismology \citep{Rui2025}. 

\citet{Fontaine1973} predicted that surface convection zones generate kG magnetic fields by a dynamo mechanism \citep[see also][]{Markiel1994,Thomas1995}. Motivated by the recent interest in such weak fields, we revisited the surface dynamo theory systematically using \textsc{mesa} stellar evolution computations, as well as a simpler analytical model based on the \citet{Mestel1952} cooling theory. We found that magnetic fields reach a maximum of several kG (for a hydrogen atmosphere) shortly following the appearance of a convection zone at a cooling time $t=t_{\rm conv}$. After that, the magnetic field gradually declines as $B\propto T_{\rm eff}\propto t^{-7/20}$, until the convective envelope couples to the degenerate core at $t=t_{\rm coup}$. This scaling relation refines the $B\propto T_{\rm eff}^{4/3}$ result of \citet{Fontaine1973}, who neglected the expansion of the convection zone. 

We compared the appearance of surface convection $t_{\rm conv}\propto M^{25/21}$ to the onset of core crystallization $t_{\rm cryst}\propto M^{-5/3}$ \citep{BlatmanGinzburg2024}, finding that for $0.5-0.9\,\msun$ white dwarfs, $t_{\rm conv}<t_{\rm cryst}<t_{\rm coup}$. We therefore always expect a phase of weak magnetization by a surface dynamo, before the convection zone is overrun by a potentially stronger magnetic field emanating from the white dwarf's crystallizing core. When comparing to observations, we found that most of the measured kG magnetic fields are in white dwarfs that have not developed a convection zone yet ($t<t_{\rm conv}$), requiring a different explanation, such as a fossil field from a previous stage of stellar evolution. We find two candidates for carbon--oxygen white dwarfs magnetized by a surface dynamo -- \phone \citep{AznarCuadrado2004} and \another \citep{Landstreet2012,Farihi2018}. Both their longitudinal magnetic fields $B_z$ and their cooling ages $t$ are consistent with our theoretical expectations. Specifically, although $t>t_{\rm cryst}$ for \phone, we find that its cooling time is shorter than the breakout time of magnetic fields from a crystallization dynamo to the surface such that $t<t_{\rm break}$ \citep{BlatmanGinzburg2024}. While high-resolution spectra probably indicate much stronger total magnetic fields $B\gg B_z$ \citep{Koester1998,Koester2009,Landstreet2017} -- ruling out a surface dynamo origin -- our candidates demonstrate the potential of spectropolarimetry to detect such weak dynamos. 

We conclude that surface convection dynamos generating kG magnetic fields are a likely by-product of white dwarf cooling. Such dynamos operate for a period $\Delta t\approx t_{\rm cryst}-t_{\rm conv}$ lasting typically for about a Gyr, but shortens for more massive white dwarfs. The small size of the convective eddies (limited by the thickness of the convection zone $\Delta r_{\rm conv}$) compared to the radius $R$ may reverse the polarity on a small spatial scale and thus frustrate the direct detection of magnetic fields using polarimetry \citep{Thomas1995}. 
None the less, our results may be useful for understanding white dwarf accretion \citep{Metzger2012,HarringtonGaraud2019,Fraser2024} and seismology \citep{Rui2025}. In addition, the convective envelopes of white dwarfs are similar in depth $\Delta r_{\rm conv}/R\lesssim 10^{-2}$ and Rossby number $\textrm{Ro}\sim 10^1$ (assuming a typical rotation period of about a day) to some magnetic F stars \citep{Mathur2014,Seach2020} -- providing an opportunity to understand the nature of the dynamo in both types of stars. 



\section*{Acknowledgements}

We thank Eliot Quataert for helpful discussions, and the anonymous reviewer for constructive comments which improved the paper. We are grateful for support from the United States -- Israel Binational Science Foundation (BSF; grant No. 2022175). RY and SG are also supported by the Israel Ministry of Innovation, Science, and Technology (grant No. 1001572596), the Israel Science Foundation (ISF; grants No. 1600/24 and 1965/24), and the German -- Israeli Foundation for Scientific Research and Development (GIF; grant No. I-1567-303.5-2024). NZR acknowledges support from the National Science Foundation Graduate Research Fellowship under grant No. DGE‐1745301.
Our study benefited from the Montreal White Dwarf Database\footnote{\url{https://www.montrealwhitedwarfdatabase.org}} \citep{Dufour2017}.

\section*{Data Availability}

The data underlying this article will be shared on reasonable request to the corresponding author.



\bibliographystyle{mnras}
\input{surface.bbl}



\appendix

\section{Helium atmospheres}\label{sec:helium}

\begin{figure}
    \centering
    \includegraphics[width=1\linewidth]{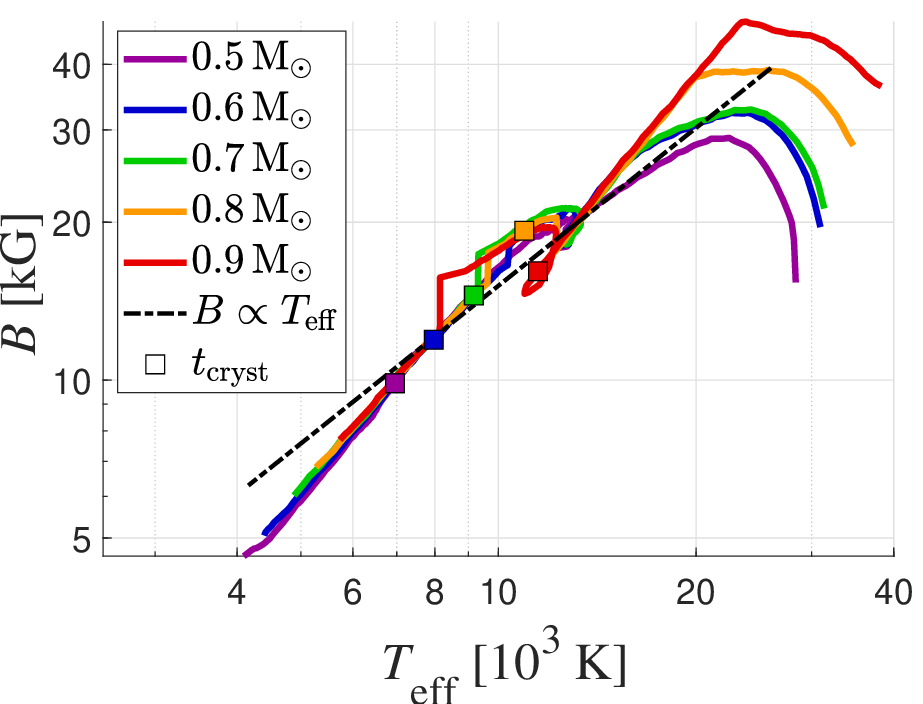}
    \caption{Similar to Fig. \ref{fig:B_Teff}, but for helium-atmosphere (DB) white dwarfs.}
    \label{fig:B_Teff_DB}
\end{figure}

While most of the paper focuses on DA white dwarfs, we also computed a set of DB white dwarfs by replacing our hydrogen atmospheres with helium using the \texttt{replace\_initial\_element} procedure; the results are presented in Fig. \ref{fig:B_Teff_DB}. The major difference is the helium's higher ionization temperature, such that a convection zone forms when $T_{\rm eff}\approx 3\times 10^4\textrm{ K}$ \citep[corresponding to the DBV instability strip, see][]{FontaineBassard2008,WingetKepler2008}. Additionally, the maximal magnetic field is higher $B_{\rm max}\approx 3\times 10^4\textrm{ G}$, as found by \citet{Fontaine1973}.

Fig. \ref{fig:B_Teff_DB} demonstrates that our main conclusions remain valid for DB white dwarfs as well. Specifically, there is always a period of an active surface dynamo prior to the onset of crystallization. At high masses, this period is actually longer for DB white dwarfs, despite their slightly earlier crystallization. While the qualitative picture is similar, some details are different for DB white dwarfs, like the deviation from our analytical $B\propto T_{\rm eff}$ power law. These differences should be studied in greater detail in the future. In particular, the non-monotonic behaviour and jumps at low $T_{\rm eff}$ are probably numerical artifacts caused by the atmospheric boundary conditions in \textsc{mesa}. 
In addition, our artificial initial conditions are not fully forgotten by the onset of convection (especially for high $M$). While they do not affect the dynamo's long decline (the focus of the current paper), these initial conditions should be improved in order to study the dynamo's role during the shorter DBV instability strip. 


\bsp	
\label{lastpage}
\end{document}